\documentclass[conference,a4paper]{IEEEtran}

\usepackage{xcolor}
\usepackage{balance}
\usepackage{multirow}
\usepackage{booktabs}
\usepackage{cite,amssymb}
\usepackage[nolist]{acronym}
\usepackage[cmex10]{amsmath}
\usepackage{subfigure}
\usepackage{color}
\usepackage{multirow}
\usepackage[normalem]{ulem}
\usepackage{soul}
\usepackage{relsize}
\usepackage{hhline}
\usepackage{amsmath}
\usepackage{algorithm}
\usepackage{algpseudocode}
\makeatletter
\def\BState{\State\hskip-\ALG@thistlm}
\makeatother

\ifCLASSINFOpdf
  \usepackage[pdftex]{graphicx}
\else
  \usepackage[dvips]{graphicx}
\fi

\ifCLASSOPTIONcompsoc
 \usepackage[caption=false,font=normalsize,labelfont=sf,textfont=sf]{subfig}
\else
 \usepackage[caption=false,font=footnotesize]{subfig}
\fi

\begin{acronym} 
\acro{5G}{fifth-generation} 
\acro{ADAS}{advanced driver assistance systems}
\acro{API}{application programming interface}
\acro{AWGN}{additive white Gaussian noise}
\acro{BEP}{beacon error probability}
\acro{BP}{beacon periodicity}
\acro{BR}{beacon resource}
\acro{BSM}{basic safety message}
\acro{BSS}{basic service set}
\acro{C-ITS}{cooperative intelligent transport systems}
\acro{C-V2X}{cellular-vehicle-to-anything}
\acro{CA}{cooperative awareness}
\acro{CAM}{cooperative awareness message}
\acro{CAM-R}{\ac{CAM} resource}
\acro{CAV}{connected and autonomous vehicle}
\acro{CCA}{clear channel assessment}
\acro{CCDF}{complementary cumulative distribution function}
\acro{CD}{collision detection}
\acro{CDF}{cumulative distribution function}
\acro{c.f.}{characteristic function}
\acro{CSI}{channel state information}
\acro{CSMA/CA}{carrier sense multiple access with collision avoidance}
\acro{D2D}{device-to-device}
\acro{DSRC}{dedicated short range communication}  
\acro{ERP}{equivalent radiated power}
\acro{FCD}{floating car data}
\acro{FD}{full duplex}
\acro{FDD}{frequency division duplex}
\acro{GPS}{global positioning system}
\acro{HD}{half duplex}
\acro{HIL}{hardware-in-the-loop}
\acro{ICRW}{intersection collision risk warning}
\acro{ICWS}{intersection collision warning system}
\acro{ITS}{intelligent transport system}
\acro{ITS-S}{intelligent transport system station}
\acro{i.i.d.}{independent identically distributed}
\acro{JDBC}{Java database connectivity}
\acro{LOS}{line of sight}
\acro{LTE}{long term evolution}  
\acro{LTE-D2D}{\ac{LTE} with device-to-device communications}  
\acro{LTE-V2V}{\ac{LTE}-vehicle-to-vehicle}  
\acro{LTE-V2X}{\ac{LTE}-vehicle-to-anything}  
\acro{MAC}{medium access control}
\acro{MCS}{modulation and coding scheme}
\acro{MD}{allocation with maximum reuse distance}
\acro{NHTSA}{National Highway Traffic Safety Administration}
\acro{NLOS}{non line of sight}
\acro{OBC}{on-board computer}
\acro{OBU}{on-board unit}
\acro{OCB}{outside the context of a BSS}
\acro{OFDM}{orthogonal frequency division multiplexing}
\acro{PDF}{probability density function}
\acro{PEP}{pairwise error probability} 
\acro{PER}{packet error rate} 
\acro{PHY}{physical}
\acro{PMF}{probability mass function}
\acro{PPP}{Poisson point process}
\acro{ProSe}{proximity services}
\acro{PRP}{packet reception probability}
\acro{QoS}{quality of service}
\acro{r.v.}{random variable}
\acro{RF}{radio frequency} 
\acro{RB}{resource block}
\acro{RR}{random resource allocation}
\acro{RSU}{road-side unit} 
\acro{SC-FDMA}{single carrier-frequency division multiple access}
\acro{SHINE}{simulation platform for heterogeneous interworking networks}
\acro{SI}{self-interference}
\acro{SINR}{signal to noise and interference ratio}
\acro{SNR}{signal to noise ratio}
\acro{SPS}{semi-persistent scheduling}
\acro{SUMO}{simulation of urban mobility}
\acro{TraaS}{TraCI as a Service}
\acro{TraCI}{traffic control interface}
\acro{TRUDI}{testing environment for vehicular applications running with devices in the loop}
\acro{SV}{smart vehicle}
\acro{TDD}{time division duplex}
\acro{TraCI}{traffic control interface}
\acro{V2C}{vehicle-to-cellular} 
\acro{V2I}{vehicle-to-infrastructure} 
\acro{V2R}{vehicle-to-roadside}
\acro{V2V}{vehicle-to-vehicle} 
\acro{V2X}{vehicle-to-anything} 
\acro{VANETs}{vehicular ad hoc networks} 

\acro{SDR}{Software Defined Radio}
\end{acronym}

\begin{document}
%
\title{A Hardware-in-the-Loop Evaluation of the Impact of the V2X Channel on the Traffic-Safety Versus Efficiency Trade-offs}

\author{\IEEEauthorblockN{
Alessandro Bazzi\IEEEauthorrefmark{1},   
Thomas Blazek\IEEEauthorrefmark{2},   
Michele Menarini\IEEEauthorrefmark{3},    
Barbara M. Masini\IEEEauthorrefmark{3}\\      
Alberto Zanella\IEEEauthorrefmark{3},       Christoph Mecklenbr\"{a}uker\IEEEauthorrefmark{2},      
Golsa Ghiaasi\IEEEauthorrefmark{4}      }                                     
\IEEEauthorblockA{\IEEEauthorrefmark{1}
\textit{University of Bologna, Italy, alessandro.bazzi@unibo.it}}
\IEEEauthorblockA{\IEEEauthorrefmark{2}
\textit{TU Wien, Austria, thomas.blazek@tuwien.ac.at, cfm@nt.tuwien.ac.at}}
\IEEEauthorblockA{\IEEEauthorrefmark{3}
\textit{CNR-IEIIT, Italy, \{michele.menarini,barbara.masini,alberto.zanella\}@ieiit.cnr.it}}
\IEEEauthorblockA{\IEEEauthorrefmark{4}
\textit{\textit{NTNU}, Norway, \ golsa.ghiaasi@ntnu.no}}  
\vspace*{-0.8cm}
}



\maketitle


\begin{abstract}\footnote{© 2020 IEEE.  Personal use of this material is permitted.  Permission from IEEE must be obtained for all other uses, in any current or future media, including reprinting/republishing this material for advertising or promotional purposes, creating new collective works, for resale or redistribution to servers or lists, or reuse of any copyrighted component of this work in other works.\\Accepted version, to be presented at EuCAP 2020. }
Vehicles are increasingly becoming connected and short-range wireless communications promise to introduce a radical change in the drivers' behaviors. Among the main use cases, the intersection management is surely one of those that could mostly impact on both traffic safety and efficiency. 
In this work, we consider an intersection collision warning application and exploit an hardware-in-the-loop (HIL) platform to verify the impact on the risk of accidents as well as the average time to travel a given distance. Besides including real ITS-G5 compliant message exchanges, the platform also includes a channel emulator with real signals. Results show that the risk of collisions can be drastically reduced, with an overall trade-off between safety and traffic efficiency. At the same time, it is shown that the presence of real channel conditions cannot guarantee the same condition of zero-risk as with ideal channel propagation, remarking the importance of channel conditions and signal processing.

\end{abstract}

\vskip0.5\baselineskip

%

\section{Introduction}
All future vehicles are expected to be equipped with wireless technologies enabling to share real time information and concurrently improve safety, optimize traffic, and provide novel services to the passengers. Probably, the process will ultimately lead to vehicles that will mostly be autonomous and connected to each other. A major step in this direction, which has long been promised, although it has not happened yet, is the full connectivity among vehicles through the use of short-range wireless communications. While various technologies are being considered and validated, including those around IEEE 802.11p and those in the area of cellular systems \cite{bazzi2019survey}, still more investigations to quantify the impact of such systems on real large-scale scenarios are being carried out.

Among the various use cases where short-range communications are expected to play a crucial role, there is what ETSI calls \ac{ICRW} \cite{ETSI_TS_101_539}. The prevention of accidents at intersections is in fact of utmost importance towards vision zero, as demonstrated by several studies (for example, they accounted for more than 20\% of crashes in Europe in \cite{AccidentReportEU18} and for about 40\% in the US in \cite{USDOT811366}). 

This work focuses on the impact of an \ac{ICRW} implementation, assuming vehicles connected using ITS-G5, which is the European technology based on IEEE 802.11p \cite{Str:J11}. Particular emphasis will be given on the one hand on the trade-off between the granted safety level and the deriving traffic efficiency, and on the other hand on the impact of a realistic signal propagation.

To this aim, we have developed an \ac{HIL} simulation platform that includes both ITS-G5 devices and a channel emulator. Differently from most of studies, which are based on simulators such as VEINS \cite{SomGerDre:J11} or iTETRIS \cite{KumLinKraHriEtal:C10} and unavoidably approximate the physical layer, in this study the signal propagation and thus the effectiveness of communications is assessed through ITS-G5 transceivers and a channel emulator. Interestingly, the platform is implemented through a multi-laboratory approach, with a core part based on the traffic simulator SUMO \cite{SUMO2012} and the \ac{TRUDI} \cite{MenMarCecBazMasZan:C19}, running in Bologna, Italy, and the signal generation and channel emulation running in Vienna, Austria.

As a reference scenario we have considered an urban area of the city of Bologna, and the performance has been evaluated in terms of both the occurred collisions (network performance) and the average travel time (efficiency of the application). The results with the implemented \ac{ICRW} application and the emulated channels are then compared with ideal reference conditions, remarking that the assumption of ideal channel conditions is not suitable to assess the real impact of vehicular applications.

\section{Use Case Definition}

This study considers connected, human driven, vehicles in an urban area with moderate traffic. Each vehicle is equipped with an \ac{ICRW} application helping the  driver to be aware of approaching vehicles with the right of way. 
The aim is to focus on what would be possible today if all vehicles were equipped with the ITS-G5 technology and \ac{ICRW} was implemented.

	\floatname{algorithm}{Algorithm}
\begin{algorithm}[t]
	\caption{Intersection collision risk warning Algorithm}\label{ICWAlgorithm}
	\begin{algorithmic}[1]
		\While {Approaching an intersection without the right of way}
		\State def $v$ : vehicle which runs ICW
		\State def $n$ : vehicle closest to the intersection among those on the other roads
		\State def $TT_{x}$: time to reach the intersection of vehicle $x$
		\State def $\Delta T_{w} [s] $: warning time threshold
		\State def $\Delta T_{A} [s] $: alarm time threshold
		\State def $TB_{x} [s] $: time to comfortably brake before the intersection
		\State def $RT_x [s] $: reaction time
		\State
		\State Constantly calculate $TT_v$ and $TT_n$
				\If {$|TT_v - TT_n|>\Delta T_{w}$}
				\State {IDLE} 
				\ElsIf {$|TT_v - TT_n|>\Delta T_{A}$}
				\State {WARNING} 
				\ElsIf {$|TT_v - TB_v - RT_v|>0$}
				\State {WARNING} 
				\Else 
				\State {ALARM} 
				\EndIf
\EndWhile
	\end{algorithmic}
\end{algorithm}  

\subsection{Intersection collision warning application}

The implemented \ac{ICRW} is summarized in Algorithm~\ref{ICWAlgorithm}. It applies to those vehicles that are approaching an intersection and do not have the right of way. Specifically, the application calculates the level of risk, based on the real time information exchanged through the so-called \acp{CAM}, which are periodic messages transmitted via ITS-G5. If necessary, it warns the driver of a dangerous situation. Although here we assume cars driven by humans, the algorithm could be easily adapted to partially or fully automated vehicles.

The level of risk is evaluated by calculating the absolute difference between the time, $TT_v$,  necessary to the vehicle 
to reach the intersection and the time, $TT_n$, necessary  to the first of the vehicles approaching from the other roads to reach the same intersection and comparing it with some given thresholds. 

More specifically, two thresholds denoted as \textit{warning threshold} $\Delta T_w>0$ and \textit{alarm threshold} $\Delta T_A<\Delta T_w$ are assumed. The variable to be compared is obtained as $\Delta TT \triangleq |TT_v-TT_n|$, where $TT_x \triangleq d_x / v_x$, $x$ stays for either $v$ or $n$, $d_x$ is the distance between the vehicle and the intersection, and $v_x$ is the speed of $x$.  

If $\Delta TT$ is below $\Delta T_w$ and above $\Delta T_A$, then the application sends a warning message to the driver. If $\Delta TT$ is below $\Delta T_A$, then it also sends a warning message if the time is sufficient for a safe and comfort break, otherwise it generates an alarm to request a sudden action by the driver.

A field trial example of an early version of the application, not including the risk levels, was presented in \cite{MenMarCecBazMasZan:C19}.

\subsection{Channel models}\label{subsec:channels}
The channel models combine pathloss and small-scale fading. The pathloss is given by
\begin{equation}
    \text{PL}(d) = L_0 + \beta10\log_{10}\left(\tfrac{d}{1m}\right)
\end{equation}
with $L_0=47.86\,$dB and the pathloss exponent $\beta=2.5$, similar to \cite{BazMasZanThi:J17}. On top of this pathloss, we model small-scale fading according to a stationary tapped-delay line model 
\begin{equation}
    h(\tau, t) = \sum_{i=1}^N10^{\tfrac{\eta_i^2}{10}}h_i(t)\delta(\tau - \tau_i).
\end{equation}
Here, $\eta^2$ is the relative path power in dB, $\tau_i$ is the relative path delay of the $i$th tap, and $h_i(t)$ is a stationary fading trace.
We use the channel model parameters for urban \ac{LOS} and \ac{NLOS} as shown Tab.~\ref{tab:chan_mods} and defined in \cite{blazek2017vehicular}. The models use $N=4$ taps, and assume the first tap to be completely static. The other taps are all Rayleigh distributed, and use half-bathtub spectra (HalfBT). $f_D$ indicates the maximum Doppler spread, while the sign shows whether the left of right side is nonzero.
\begin{table}[!t]
  \caption{Link Level Channel Models \cite{blazek2017vehicular}}
  \label{tab:chan_mods}
  \centering
  \begin{tabular}{lcrrrc}
    \toprule
Name & Tap & $\eta_i^2$ [dB] & $\tau_i$ [ns] & $f_{i,d}$ [Hz] & Profile \\
    \midrule
    \multirow{4}{*}{ \parbox[t]{1.5cm}{Urban\\LOS}} & $i=1$ & $0$&$0$ & $0$& Static\\   
 & $i=2$ & $-8$ & $117$ & 236 & HalfBT\\ 
 & $i=3$ & $-10 $ &$183$  & $-157$& HalfBT\\ 
 & $i=4$ & $-15$& $333$ & $492$ & HalfBT \\
    \midrule
    \multirow{4}{*}{ \parbox[t]{1.5cm}{Urban\\NLOS}} & $i=1$ & $0$ &$0$ & $0$& Static \\ 
 & $i=2$ &$-3$ &$267$ & $295$& HalfBT \\ 
 & $i=3$ & $-4$&$400$ & $-98$& HalfBT \\   
 & $i=4$ & $-10$   & $533$ &$591$ & HalfBT \\
    \bottomrule
  \end{tabular}
\end{table}
\subsection{Output metrics}\label{subsec:metrics}

The following two output metrics are used, one dealing with safety and the other with traffic efficiency. For both of them, the reference situation is when all drivers are careful and respect the rights of way, in that case no collision occurs.

\subsubsection{Average collisions per hour} It indicates the number of collisions between two vehicles that have occurred during the simulations, averaged over 1~hour. This metric highlights the effectiveness of the \ac{ICRW} to improve safety at intersections.

\subsubsection{Time improvement to careful case} Considering the time needed to travel a reference distance (e.g., 1\,km), it is defined as the time difference between the time spent by the vehicle to travel the distance and the time that the vehicle would spend 
if all drivers were always careful and diligently respected the rights of way. This metric  quantifies the improvement in traffic efficiency and a negative value means that more time is (on average) required to cover the same distance.

\begin{figure*} [t]
	\centering
	\includegraphics[trim={100 100 100 100},width=0.62\linewidth,draft=false]{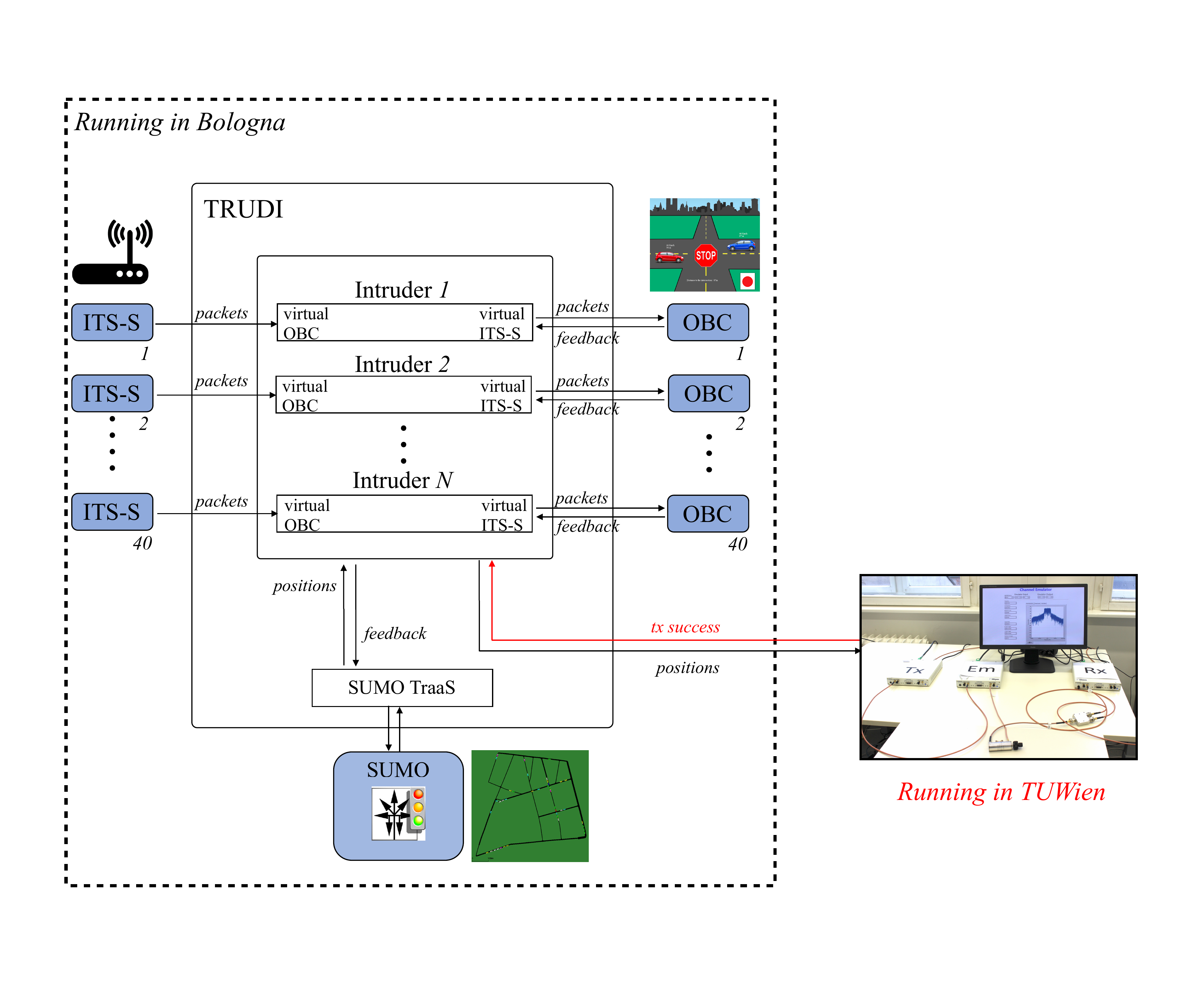}
	\caption{Structure of the HIL platform.}
	\label{Fig:Structure} 
\end{figure*}

\section{The Hardware-in-the-Loop Platform}

The \ac{HIL} platform consists of two main parts, one based on the HIL platform TRUDI \cite{MenMarCecBazMasZan:C19} and the traffic simulator SUMO \cite{SUMO2012}, running in Bologna, and the other focusing on signal generations and channel emulation, running in Vienna. A block scheme of the platform is shown in Fig.~\ref{Fig:Structure}. All simulations run in real time, with a one-to-one correspondence between simulated and simulation time.

\subsection{TRUDI}

The TRUDI platform, detailed in \cite{MenMarCecBazMasZan:C19}, is based on the so-called intruders, which act between the wireless devices (or \acp{ITS-S}) and the corresponding controllers running inside the cars (or \acp{OBC}). Each intruder mimics the OBC for the ITS-S and the ITS-S for the OBC, thus receiving and passing all exchanged messages, with the possibility to alter the content and simulate some events. In particular, the intruders modify the GPS coordinates received from the ITS-S with those obtained from the traffic simulator SUMO in order to reproduce the mobility on the road.
In addition, it discards some of the messages to reproduce the transmission errors due to channel impairments. 

At the same time, TRUDI receives from the OBCs a feedback indicating warnings or alarms from the ICRW application, which are then used to modify the drivers' behavior in the traffic simulator.

\subsection{SUMO and the behavior of drivers}\label{subsec:SUMO}

SUMO \cite{SUMO2012} is an open source microscopic and continuous traffic simulator. Among the several features of SUMO, it is also possible to modify the behavior of the drivers, for example making them more aggressive. In order to investigate the impact of the ICRW application, in this work we have exploited a feature of SUMO that allows to set the drivers to either respect or not the right of way at intersections. 

First of all, we consider the following benchmark.
\begin{itemize} 
\item \textit{Careful case}: all drivers are always careful and respect the correct rights; this case corresponds to the basic operation of SUMO.
\end{itemize}
Figures will not explicitly show results regarding this case. In fact, looking at safety it implies no collisions by design and, focusing on traffic efficiency, the adopted metric is obtained as an improvement compared to this case.

As a further reference, we also evaluate the following case.
\begin{itemize} 
\item \textit{No-App} case: all drivers are always distracted (i.e., they do not respect the right of way). 
\end{itemize}

All other cases imply the exploitation of wireless communications and the implemented application, as hereafter detailed.
\begin{itemize} 
\item \textit{ICRW cases}: all drivers are normally distracted, but become careful when alerted by the application; more specifically, if the ICRW warning is activated, then the driver becomes careful with i.i.d. probability 0.5; if the ICRW alarm is triggered, then the driver becomes careful with probability 1 (i.e., they always become careful); the driver is again distracted once the vehicle crosses the intersection.
\end{itemize}
Please note that, for the sake of readability, the acronym ICRW is left implicit in the figures of Section~\ref{sec:results} and all results except for the \textit{No-App} belong to these cases.

\subsection{Channel emulation}
The testbed for emulation consists of two \ac{SDR} platforms, namely, two NI USRP-2953R modules each equipped with 40 MHz of operation and two RF channels. One USRP is running the NI 802.11 project code which was modified to support 802.11p stack. The  \ac{SDR} is set up in RF loop-back configuration, with channel 0 acting as transmitter, and channel 1 acting as the receiver. 
The other USRP module programmed as channel emulator is placed between the transmitter and the receiver. The \ac{SDR} implements the time-variant channel emulator described in \cite{golsa2018emulator} and replicates the channel described in Tab.~\ref{tab:chan_mods}. 

At each periodic generation of the \acp{CAM}, a vector with the position of all the vehicles in the scenario is sent from TRUDI to the channel emulator.  Based on this information, the first device logs transmission successes and failures at the receiving side, and records these events. A matrix with a map of correctly received and lost messages is returned to TRUDI. Currently, a limited number of vehicles is considered, thus interference and collisions are assumed negligible.

\section{Results}\label{sec:results}

\begin{figure} [t]
	\centering
	\includegraphics[width=0.70\linewidth,draft=false]{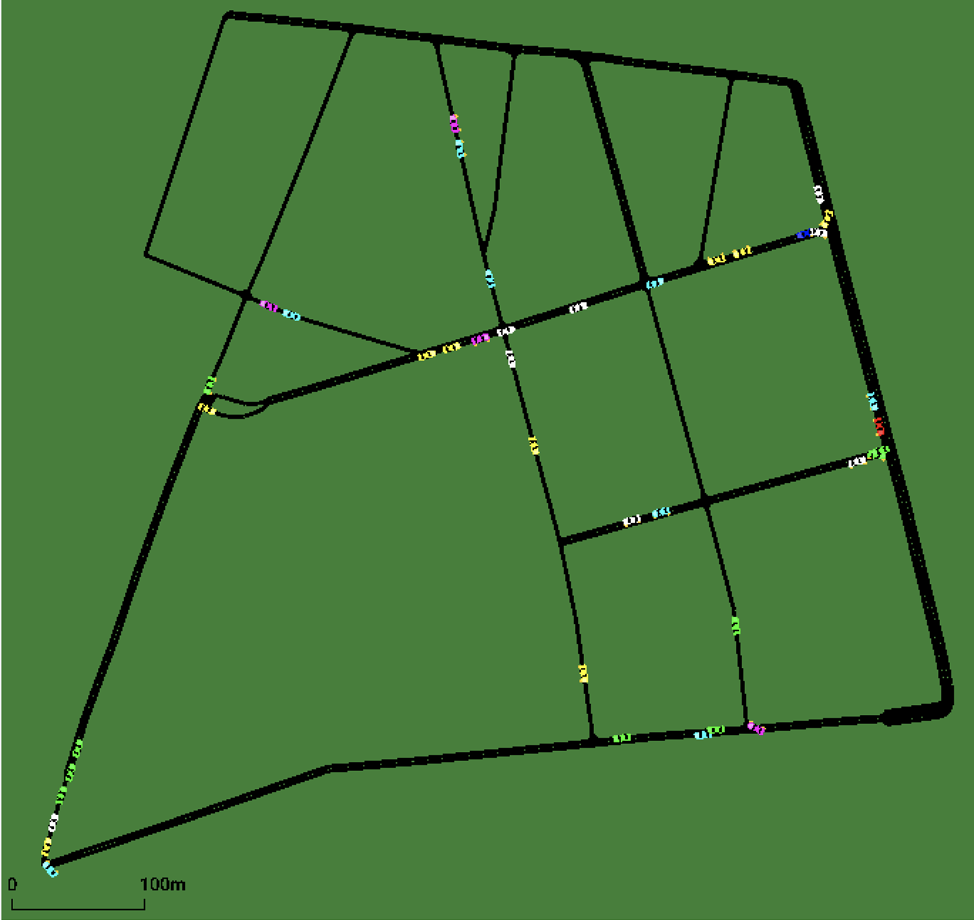}
	\caption{Example snapshot of the scenario simulated in SUMO.}
	\label{Fig:Snapshot} 
\end{figure}

\begin{figure*} [t]
	\centering
	\subfigure[Average collisions per hour.]{
		\includegraphics[width=0.45\linewidth,draft=false]{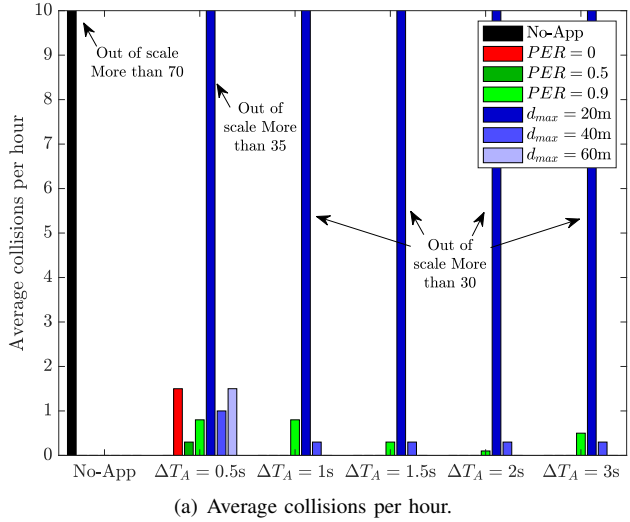}}~~~~~
	\label{fig:RefTime}\subfigure[Time difference compared to the case where drivers are careful.]{
		\includegraphics[width=0.45\linewidth,draft=false]{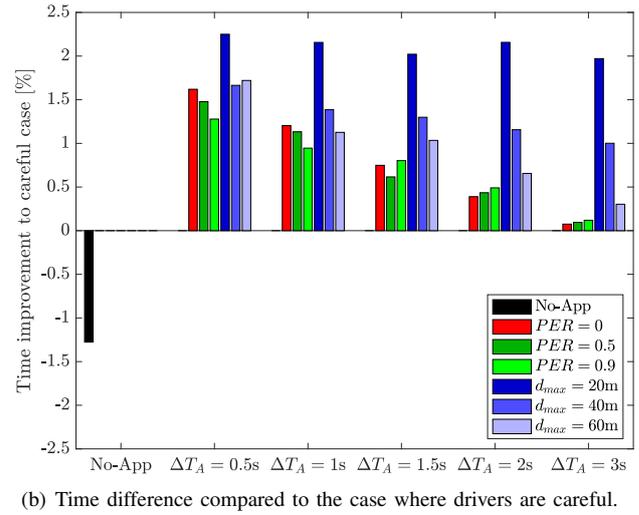}}
	\caption{Results with reference channels. The No-App case is compared to: 1) packet losses with fixed probability, independent to the distance ($PER=P_X$, where $P_X$ is the packet error rate); 2) threshold source-destination distance ($d_{max}=d_X$), below which packets are correct and above which  are lost.  }
	\label{fig:Ref}
\end{figure*}

\begin{figure*} [t]
	\centering
	\label{fig:EmuColl}\subfigure[Average collisions per hour.]{
		\includegraphics[width=0.45\linewidth,draft=false]{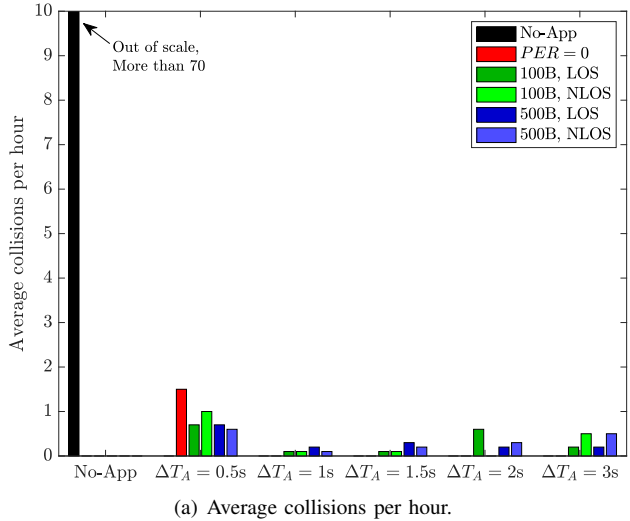}}~~~~~
	\label{fig:EmuTime}\subfigure[Time difference compared to the case where drivers are careful.]{
		\includegraphics[width=0.45\linewidth,draft=false]{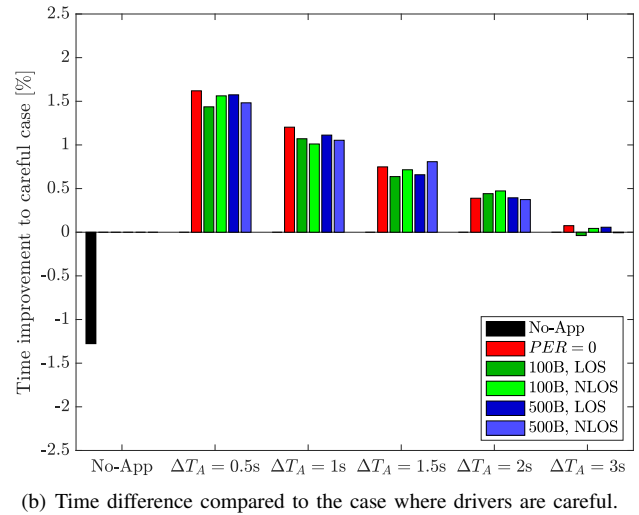}}
	\caption{Results with emulated channels. The No-App case is compared to the ICRW cases with ideal channel (no packet losses, $PER=0$) and the four emulated channels: 100 byte packets, LOS and NLOS conditions, and 500 byte packets, LOS and NLOS conditions.}
	\label{fig:Emu}
\end{figure*}

Results obtained in the scenario are exemplified in Fig.~\ref{Fig:Snapshot}. It corresponds to a portion of the city of Bologna derived from OpenStreetMap~(\cite{OpenStreetMap}) in the premises of the Engineering School of the University of Bologna. Each simulation consists of 10 hours with 40 vehicles moving with random directions and a maximum speed of 50\,km/h. The main settings are also summarized in Table~\ref{Tab:settings}.
Two groups of results, assuming either ideal or realistic channel conditions, are shown in Figs.~\ref{fig:Ref} and~\ref{fig:Emu}, respectively. The two metrics detailed in Section~\ref{subsec:metrics} are shown varying the alarm threshold $\Delta T_A$. The warning  threshold is always set to $2 \cdot \Delta T_A$ and left implicit in the following.

\begin{table}
\caption{Main settings
\vspace{-2mm}
\label{Tab:settings}}
\centering
\begin{tabular}{p{4.5cm}p{1.2cm}}
		\hline
\textbf{Scenario} & \\
	Simulated time & 10 hours \\
	Number of vehicles & 40 \\
	Maximum speed of vehicles & 50 km/h \\
	\hline
\textbf{Algorithm} & \\
    Warning time threshold $\Delta T_w$ & $2 \cdot \Delta T_A$ \\
    Alarm time threshold $\Delta T_A$ & Variable \\
    Deceleration (for the time to break $TB$) & 4 $m/s^2$\\ 
    Reaction time $RT$ & 1~s \\
    \hline
\end{tabular}
\end{table}


Fig.~\ref{fig:Ref} compares the No-App case (see Section~\ref{subsec:SUMO}) with the ICRW cases, assuming ideal channels. Specifically, the following channels are considered:
\begin{enumerate}
    \item $PER = P_X$: all packets are lost with i.i.d. probability $P_X$, independently to the position of the vehicles;
    \item $d_{max} = d_X$m: all packets exchanged between vehicles whose distance is less than $d_X$~m are correctly received, otherwise they are lost.
\end{enumerate}

It can be noted from Fig.~\ref{fig:Ref}(a) that a lower number of collisions is always granted by the application. The reduction is more or less halved if the coverage is limited to 20\,m ($d_{max}=20$\,m) and drastically reduced in all the other cases. It can also be observed that a value of $\Delta T_A=0.5$\,s is insufficient to avoid all the collisions, neither with an ideal channel, whereas a value of $\Delta T_A=1$\,s allows to have no accidents when  the error rate is kept at 50$\,\%$ or the range sufficiently large ($d_{max}=60$\,m). 

Looking at Fig.~\ref{fig:Ref}(b), a positive impact on the travel time can also be noted, except for the No-App case. This means that the vehicles reduce their speed only when strictly needed, improving  the traffic efficiency. As expected, the time improvement reduces with an increase of $\Delta T_A$, making clear  the trade-off between traffic safety and efficiency. It is worth noting that the shown improvement, which might appear small, is obtained considering the entire route of the vehicles, even if the application is effective only near the intersections. 

Fig.~\ref{fig:Emu} shows the same outputs when the realistic channels detailed in Section~\ref{subsec:channels} are used. Again, a drastic reduction of accidents (Fig.~\ref{fig:Emu}(a)) and an improvement of traffic efficiency (Fig.~\ref{fig:Emu}(b)) can be observed, with a trade-off between them that depends on the chosen value of $\Delta T_A$. 

One important aspect to remark comparing Figs.~\ref{fig:Ref} and~\ref{fig:Emu} is that with realistic channel conditions the number of collisions occurred in a simulation never goes to zero, neither fixing $\Delta T_A=3$\,s and basically there is no advantage in terms of traffic efficiency.  This is due to the fact that in a real scenario errors tend to be strongly correlated, and bursts of errors occur, causing the presence of periods of unawareness even when the average error rate is limited. 

\section{Conclusion}

In this work, we have considered an \ac{ICRW} application used in an urban scenario to improve safety and traffic efficiency, with the aim to investigate the impact of a realistic channel. The objective has been achieved exploiting a \ac{HIL} platform where real ITS-G5 signals are exchanged through a channel emulator configured with realistic channels. Results have indeed demonstrated a certain trade-off between traffic safety and efficiency. In addition, the impact of realistic channels causes residual accidents in all situations, which were not observed for ideal channel conditions.

\section*{Acknowledgment}
\footnotesize
This work has been partially conducted within the DARVIS project funded by the Austrian Aeronautics Research and Technology Program TAKEOFF under grant agreement No 867400.

\bibliographystyle{IEEEtran}
\bibliography{biblio.bib}

\end{document}